\documentclass[10pt,letterpaper,twocolumn]{article} 

\usepackage{ol2}
\usepackage[draft,implicit=false]{hyperref}
\usepackage{amsmath}

\begin{document}

\twocolumn[ 

\title{Measurement of the squeezed vacuum state by a bichromatic local oscillator}


\author{Wei Li,$^{1,2}$, Xudong Yu,$^{1,2}$ and Jing Zhang$^{1,3,*}$}

\address{$^{1}$The State Key Laboratory of Quantum Optics and Quantum Optics Devices, Institute of Opto-Electronics, \\
Shanxi University, Taiyuan 030006, P.R.China\\
$^{2}$ Collaborative Innovation Center of Extreme Optics, Shanxi
University, Taiyuan 030006, P.R.China\\
$^{3}$Synergetic Innovation
Center of Quantum Information and Quantum Physics,\\ University of
Science and Technology of China,
Hefei, Anhui 230026, P. R. China\\

$^*$Corresponding author: jzhang74@sxu.edu.cn, jzhang74@yahoo.com }

\begin{abstract}We present the experimental measurement of a squeezed vacuum state
by means of a bichromatic local oscillator (BLO). A pair of local
oscillators at $\pm$5 MHz around the central frequency $\omega_{0}$
of the fundamental field with equal power are generated by three
acousto-optic modulators and phase-locked, which are used as a BLO.
The squeezed vacuum light are detected by a phase-sensitive
balanced-homodyne detection with a BLO. The baseband signal around
$\omega_{0}$ combined with a broad squeezed field can be detected
with the sensitivity below the shot-noise limit, in which the
baseband signal is shifted to the vicinity of 5 MHz (the half of the
BLO separation). This work has the important applications in quantum
state measurement and quantum information. \end{abstract}


 ] 

\noindent Squeezed state of the light is an important resource of
the quantum information
\cite{Furusawa,Bowen,Li,Ukai,Jia,Su,Liu,Braunstein,Weedbrook} and
quantum metrology \cite{five, six, seven, eight, nine}. Especially,
in the modern research focus, the squeezed state becomes crucial for
the gravitation wave detection. In recent years, some significant
improvement have been made in this field, such as the 12.7 dB
squeezing has been obtained \cite{ten}, the very lower frequency
squeezing measurement has been realized and the frequency-dependence
squeezing has been investigated \cite{Chelkowski}. A single
broadband squeezed light can be split into N pairs of upper and
lower single sideband fields with spatial separation, which
correspond to N independent EPR entangled fields \cite{jzhang}. This
scheme was demonstrated experimentally by using a pair of
frequency-shifted local oscillators to measure this EPR entanglement
\cite{Huntington,Hage}. The theoretical scheme based on a
bichromatic local oscillator (BLO) to detect the squeezed state was
proposed \cite{Marino}, in which several advantages and applications
were given. The phase-sensitive detection with a BLO or a
double-sideband signal field were studied \cite{Fan,LiW,appel}. In
this paper, we utilize a BLO to detect a broadband squeezed light
with a phase-sensitive balanced-homodyne detection. This work
demonstrates quantum correlation between the upper and lower
sideband modes \cite{jzhang} of a single broadband squeezed light
from another perspective. Generating and measuring the low frequency
squeezing for the terrestrial gravitational wave detectors are very
tough because of the extremely challenges in the technique. The BLO
technique can circumvent the challenge of detecting low frequency
squeezing, which is usually obscured by technical noise. We present
the result that the baseband signal is shifted into the vicinity of
5 MHz ((the half of the BLO separation) and sub-shot-noise detection
is implemented. Thus this work with the BLO and broadband squeezing
can be used to enhance the signal-to-noise ratio (SNR) of an
interferometer for lower frequency phase measurement
\cite{Yurke,Zhai}.

\begin{figure}
\centerline{
\includegraphics[width=2in]{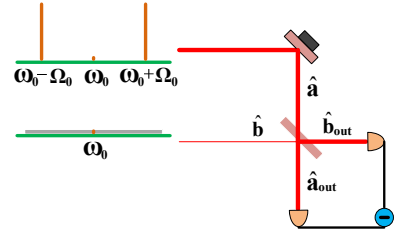}
} \vspace{0.1in}
\caption{ The schematic diagram of measuring a single broadband
squeezed light with the central frequency $\omega_{0}$ using a
phase-sensitive balanced-homodyne detection with a BLO. \label{Fig1}
}
\end{figure}

The schematic diagram of the detection is shown in Fig. 1. A strong
BLO (at $\pm\Omega_{0}$ around the central frequency $\omega_{0}$ of
the fundamental field with equal power) is mixed with the signal
light field at a 50/50 beam splitter. The relative phase $\theta$ of
the local oscillator and the signal field can be controlled by the
reflective mirror mounted on a PZT (piezoelectric transducer). The
annihilation operators of the bichromatic local oscillator and the
signal field can be written as
$\hat{a}(t)=\hat{a}_{+}(t)\exp[-i(\omega_{0}+\Omega_{0})t]+\hat{a}_{-}(t)\exp[-i(\omega_{0}-\Omega_{0})t]$
and $\hat{b}(t)=\hat{b}_{0}(t)\exp{(-i\omega_{0}{t})}$, where
$\hat{a}_{+(-)}(t)$ and $\hat{b}_{0}(t)$ are the slow varying
operators of the fields. The output fields of the 50/50 beam
splitter are
\begin{eqnarray}
\hat{a}_{out}(t)= [\hat{a}(t)e^{i\theta}+\hat{b}(t)]/\sqrt{2},
\\\hat{b}_{out}(t)=[\hat{a}(t)e^{i\theta}-\hat{b}(t)]/\sqrt{2}.
\end{eqnarray}
Therefore, the normalized difference of the photocurrents of the two
detectors may be
\begin{eqnarray}
\hat{i}(t)=\frac{1}{\sqrt{2}a}[\langle\hat{a}^{\dag}(t)\rangle\hat{b}(t)e^{-i\theta}+\langle\hat{a}(t)\rangle\hat{b}^{\dag}(t)
e^{i\theta}],
\end{eqnarray}
where the fields satisfy
$\langle\hat{a}_{+}\rangle=\langle\hat{a}_{-}\rangle=a\gg\langle\hat{b}_{0}\rangle\sim0$.
Therefore the signal field may be the vacuum state or the squeezed
vacuum state. And the bichromatic local oscillator is a pair of the
strong and equal coherent states.

\begin{figure}
\centerline{
\includegraphics[width=3.5in]{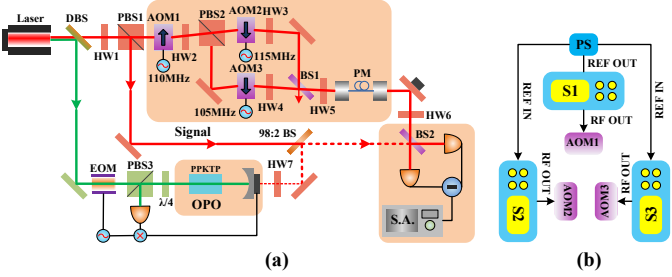}
} \vspace{0.1in}
\caption{ (a) The setup for the experimental realization of
measuring vacuum squeezing by a phase-sensitive balanced-homodyne
detection with a bichromatic local oscillator. DBS1,2: dichroic beam
splitter; HW1-4: half wave plate; BS1,2: 50/50 beam splitter;
AOM1-3: acousto-optic modulator; PM: single-mode
polarization-maintaining optical fiber. EOM: electro-optical
modulator. OPO: optical parametric oscillator. SA: spectrum
analyzer. (b) The scheme for locking the relative frequency and
phase of the double sidebands. PS: power splitter for radio
frequency. \label{Fig2} }
\end{figure}

The difference of the photocurrents analyzed at the radio frequency
$\Omega$ is expressed as
\begin{eqnarray}
\hat{i}(\Omega)&=&\frac{1}{\sqrt{2}}[\hat{b}(\Omega_{0}-\Omega)+\hat{b}(-\Omega_{0}-\Omega)]e^{-i\theta}\nonumber\\&&
+\frac{1}{\sqrt{2}}[\hat{b}^{\dag}(\Omega_{0}+\Omega)+\hat{b}^{\dag}(-\Omega_{0}+\Omega)]e^{i\theta}.
\end{eqnarray}
Here, we express the quadrature component of the signal field around
the central frequency $\omega_{0}$, which easily compare with the
measurement with a single local oscillator at $\omega_{0}$.
Therefore, the quadrature component of the signal field can be
defined as
$\hat{Q}_{S}(\nu,\theta)=\hat{b}(-\nu)e^{-i\theta}+\hat{b}^{\dag}(\nu)e^{i\theta}$
at the analysis frequency $\nu$, which includes the up and down
sideband modes $\pm\nu$. The quadrature amplitude ($\theta=0$) can
be $\hat{X}_{S}(\nu)=\hat{b}(-\nu)+\hat{b}^{\dag}(\nu)$ and the
quadrature phase ($\theta=\pi/2$)
$\hat{Y}_{S}(\nu)=-i[\hat{b}(-\nu)-\hat{b}^{\dag}(\nu)]$. The
difference of the photocurrents with the bichromatic local
oscillator (Eq. 4) will give the information of the quadrature
component of the signal field
\begin{eqnarray}
\hat{Q}_{B}(\Omega,\theta)&=&\frac{1}{\sqrt{2}}[\hat{Q}_{S}(|\Omega_{0}-\Omega|,\theta)+\hat{Q}_{S}(\Omega_{0}+\Omega,\theta)].
\end{eqnarray}
Here, $\Omega$ is the analysis frequency of the bichromatic local
oscillator detection. We can know from here that there are two pairs
of the sideband modes $\pm|\Omega_{0}-\Omega|$ and
$\pm(\Omega_{0}+\Omega)$ to be measured by the bichromatic local
oscillator. The arbitrary quadrature component of the signal field
can be measured by scanning the relative phase of $\theta$. So when
$\theta=0$, the difference of the photocurrents will give the
information of the quadrature amplitude of the signal field
$\hat{X}_{B}(\Omega)=\frac{1}{\sqrt{2}}[\hat{X}_{S}(|\Omega_{0}-\Omega|)+\hat{X}_{S}(\Omega_{0}+\Omega)]$,
and when $\theta=\pi/2$, the quadrature phase
$\hat{Y}_{B}(\Omega)=\frac{1}{\sqrt{2}}[\hat{Y}_{S}(|\Omega_{0}-\Omega|)+\hat{Y}_{S}(\Omega_{0}+\Omega)]$.

For the broad quadrature phase squeezing as the input signal field,
the quadrature components satisfy
$\langle\Delta^2\hat{Y}_{S}(|\Omega_{0}-\Omega|)\rangle=\langle\Delta^2\hat{Y}_{S}(\Omega_{0}+\Omega)\rangle=e^{-2r}<1$
and
$\langle\Delta^2\hat{X}_{S}(|\Omega_{0}-\Omega|)\rangle=\langle\Delta^2\hat{X}_{S}(\Omega_{0}+\Omega)\rangle=e^{2r}>1$,
where $r$ is the squeezing parameter. Correspondingly, we can obtain
the variance of the quadrature components of the signal field by
means of a bichromatic local oscillator,
\begin{eqnarray}
\langle\Delta^2\hat{Y}_{B}(\Omega)\rangle=e^{-2r}<1,\langle\Delta^2\hat{X}_{B}(\Omega)\rangle=e^{2r}>1.
\end{eqnarray}
So the difference of the output photocurrents at the analyzed
frequency $\Omega$ presents quadrature phase squeezing of the two
pairs of the sideband modes $\pm|\Omega_{0}-\Omega|$ and
$\pm(\Omega_{0}+\Omega)$. Thus, when the squeezing spectrum of the
input field can reach the frequency of $2\Omega_{0}$, we may obtain
the baseband signal around $\omega_{0}$ of the input field below the
shot-noise limit at the analyzed frequency $\Omega_{0}$ of the
measured noise spectrum.

\begin{figure}
\centerline{
\includegraphics[width=3.5in]{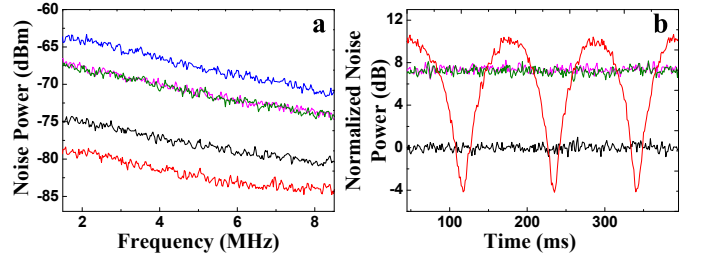}
} \vspace{0.1in}
\caption{ The measured noise power spectra. (a) The noise power
spectra at the analysis frequencies from 1.5 MHz to 8.5 MHz with the
resolution bandwidth of 100 kHz and the video bandwidth of 300 Hz.
The blue line: the anti-squeezing component; The red line: the
squeezing component; (b): The noise power spectra at the analysis
frequency of 5 MHz are measured by scanning the relative phase of
$\theta$. The red line: the quadrature component; The black line:
shot noise limit; The green and pink lines: the noise power spectrum
when the single sideband of the bichromatic local oscillator is used
respectively. Here, the electric (dark) noise is about 14 dB below
SNL. \label{Fig3} }
\end{figure}

Figure 2 shows the experimental setup. The laser source, which is a
diode-pumped external-cavity frequency doubled laser, provides the
second-harmonic light of 450 mW at 532 nm and the fundamental light
of 200 mW at 1064 nm simultaneously. The second harmonic light is
used to pump the OPO (optical parametric oscillator). The
fundamental light is separated into two parts. One is utilized as
the auxiliary beam to adjust the interference of the local
oscillator and the detected field. The other is used to generate the
bichromatic local oscillator by three acousto-optic modulators and
phase-locked technology. The frequency shifts of AOM1, AOM2 and AOM3
are +110 MHz, -115 MHz and -105 MHz respectively. The two
frequency-shifted laser beams at $\omega_{0}\pm\Omega_{0}$
($\Omega_{0}=5$ MHz) are combined on 50$\%$ BS1 with the same
polarization. The bichromatic laser field is coupled into a
single-mode polarization-maintaining fiber to filter the spatial
modes. As above discussion, only when the up- and down-shifted
frequencies ($\Omega_{+}$ and $\Omega_{-}$) around the central
frequency $\omega_{0}$ of the bichromatic local oscillator are same
($\Omega_{+}=\Omega_{-}=\Omega_{0}$) and the relative phase is
fixed, the measurement for the broad squeezed light becomes the
balanced homodyne detection. The two methods of locking the relative
phase between up- and down-shifted laser fields have been developed
in our previous work \cite{LiW}. Here we employ the clock
synchronization of the signal generators. The three signal
generators for driving the AOMs can be locked together in frequency
and phase by using the same reference (clock) frequency.

The resonator for the squeezed light is a semi-monolithic
triple-resonant OPO. The cavity is 38 mm long, and consists of a 10
mm long PPKTP crystal. The front facet of crystal acts as the input
coupler, has a transmittance of 5 $\%$ for 532 nm and is highly
reflective for 1064 nm. The rear facet is anti-reflective for both
532 nm and 1064 nm. The output mirror of the OPO has a transmittance
of 12.5 $\%$ for 1064 nm and is highly reflective for 532 nm. The
cavity bandwidth is around 70 MHz. The PPKTP is a type I
quasi-phase-matching crystal and its phase-matching condition is
achieved by the temperature controller \cite{Ye,Di}. The OPO cavity
is locked according to the PDH technique, the error signal is
derived from the reflected pump field. By adjusting the reflective
mirror of LO mounted on PZT, the relative phase $\theta$ between the
bichromatic LO and the squeezed light field is changed. The squeezed
state is mixed with the bichromatic local oscillator on the 50/50
beam splitter with the interference fringe visibility of 98 $\%$. At
last the two output fields of the beam splitter are measured by two
balanced detectors.

\begin{figure}
\centerline{
\includegraphics[width=2.5in]{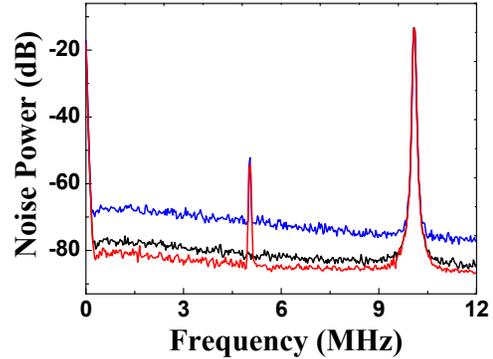}
} \vspace{0.1in}
\caption{ The noise spectra at analysis frequency from 0 MHz to 12
MHz with a baseband signal field around $\omega_{0}$. The red line:
the baseband signal peak at 5 MHz with the squeezed quadrature
component; The blue line: the baseband signal peak with the
antisqueezed quadrature component. The black line: shot noise limit.
The peak at 10 MHz is due to the beatnote of the bichromatic local
oscillator. Here, the electric (dark) noise is about 14 dB below
SNL. The power of the LO and the signal is 4 mW and about 200 nW
respectively. RBW = 30 kHz, VBW = 1 kHz, Sweep time=200 ms.
\label{Fig4} }
\end{figure}

Fig. 3 shows the noise power spectra measured by means of a
bichromatic local oscillator. The power of the pump field of the OPO
is about 40 mW. The total power of the bichromatic local oscillator
is set 4.0 mW. In the analysis frequency regime from 1.5 MHz to 8.5
MHz, the squeezing value is $4.1\pm0.2$ dB
($\langle\Delta^2\hat{Y}\rangle=0.39\pm0.02$), and the
anti-squeezing is about $10.1\pm0.2$ dB
($\langle\Delta^2\hat{X}\rangle=10.2\pm0.5$) higher than SNL (Fig.
3(a)). The noise power spectra at the analysis frequency of 5 MHz
with zero span are measured by scanning the relative phase of
$\theta$ (Fig. 3(b)). It demonstrates that the bichromatic local
oscillator is a phase-sensitive balanced-homodyne detection and the
arbitrary quadrature components can be measured (the red line in
Fig. 3(b)). When one of the bichromatic local oscillator is blocked
to become a single local oscillator, the detection is changed into
the heterodyne detection and the noise spectrum is $7.3\pm0.2$ dB
higher than the SNL, which is insensitive to the relative phase of
$\theta$ (The green and pink lines in Fig. 3). It is consistent well
with the estimation 7.25 dB
($(\langle\Delta^2\hat{Y}\rangle+\langle\Delta^2\hat{X}\rangle)/2=5.3$
) according to the above measurements.

When a baseband signal field around $\omega_{0}$ is added into the
squeezed vacuum field by a 98/2 beam splitter, the noise spectra at
analysis frequency from 0 MHz to 12 MHz is given in Fig. 4 by means
of the bichromatic local oscillator detection. The baseband signal
peak appears at 5 MHz, whose sensitivity is 4.1 dB below the
shot-noise limit when the squeezed quadrature component is detected
(the red line in Fig. 4). It demonstrates that the baseband signal
is shifted into the vicinity of 5 MHz ((the half of the BLO
separation) and sub-shot-noise detection is obtained.

In conclusion, we study a phase-sensitive balanced-homodyne
detection with a bichromatic local oscillator. The baseband signal
field around $\omega_{0}$ with a broad squeezed field is detected
and the sensitivity of the signal can be below the shot-noise limit.
This detection scheme can be employed in gravitational-wave
detection and the quantum information process.

We acknowledge the financial support from the National Basic
Research Program of China (Grant No. 2011CB921601), National Natural
Science Foundation of China (NSFC) (Grant No. 11234008, 11361161002,
61571276), Natural Science Foundation of Shanxi Province (Grant No.
2015011007).

\end{document}